\documentstyle[a4,12pt]{article}
\begin{document}
\input psbox
\large
\begin{center}
{\bf Optical follow up of the GRB 990123 source from UPSO, Nainital }

\vspace{0.5cm}

\normalsize
 {\sf R. Sagar, A. K. Pandey, V .Mohan, R. K. S. Yadav, Nilakshi } \\
{\it U. P. State Observatory, Manora Peak, Naini Tal - 263 129, India} \\
\vspace{0.5cm}
 {\sf D. Bhattacharya} \\
 {\it Raman Research Institute, Bangalore 560080, India } \\
\vspace{0.4cm}
      and   \\
\vspace{0.4cm}
{\sf A.J. Castro-Tirado} \\
{\it IAA-CSIC, P.O. Box 03004, E-18080, Granada, Spain } \\
\end{center}

\begin{abstract}
The CCD magnitudes in Johnson $BV$ and Cousins $R$ photometric passbands 
are determined for the optical transient of GRB 990123. These
observations have been taken approximately 12 hours after trigger of the 
gamma-ray burst. BVR photometric light curves are obtained by combining
 the published data with the present measurements. They indicate that
the flux decay constants in $BV$ and $R$ passbands are almost the same. Its
value is $1.1\pm 0.06$. Such decay is quite consistent with the fireball 
models for the optical transient. 
\end{abstract}

\section{\bf Introduction}
After launching  of the Italian-Dutch satellite BeppoSAX in mid-1996, it 
has  been possible to obtain position of Gamma-ray Burst (GRB) with  an 
accuracy better than 3$-$5 arc minutes within hours of occurance. The follow up 
observations of relatively long-lasting afterglows in X-ray, optical and 
radio regions have now therefore become routine. Such multi-wavelength 
observations have contributed significantly to our understanding of GRB sources.

Piro et al. (1999) reported a very bright GRB on 23 January 1999 at UT 09:46:59.
Within 3.5 hour after the burst trigger time, Odewahn et al. (1999)       
discovered a $R =$ 18th magnitude object not present in the Digital Sky Survey
that was tentatively identified as the GRB optical counterpart. Based 
on images obtained much earlier, Akerlof \& McKay (1999) discovered a 
rapidly fading object at $\alpha =  15^h 25^m 30.^s34,\delta = 
+44^{\circ} 45^{'} 59.5^{''}$ (J2000), consistent with the position of
the optical transient reported by Odewahn et al. (1999).
 Within 10 minutes, the luminosity of the optical transient varied 
from V = 11.82 to 14.53 mag after peaking at V = 8.95 mag just 
after the first exposure. These spectacular events led Djorgovski et al. 
(1999) to suggest that GRB 990123 is lensed by a foreground galaxy. This
scenario is not supported by the absorption line spectroscopic redshift 
measurements by Kelson et al. (1999) and Hjorth et al. (1999). They found that 
the value of redshift  for optical transient is 1.60. A more extended 
discussion on this topic can be seen in  Andersen et al. (1999).

We started optical observations of the optical transient at Nainital 
approximately 12 hours after the trigger of the burst under a long term 
collaborative programme between Danish, German, Indian, Italian, Mexican and 
Spanish astronomers coordinated by one of us (AJCT). Successful optical
observations have been carried out in UBVRIJK on 24 January 99, and in UBVRI 
from 25-30 January 99, at ground-based observatories (0.8-m IAC80, 1.5-m TCS,
2.2-m CAHA, 2.5-m NOT and 3.5-m CAHA) under this collaboration,
in addition to observations from the U.P. State Observatory (UPSO), Nainital. 
The photometric results of the UPSO observations are presented here. 
The decay light curves are presented in $BVR$ photometric passbands by 
combining them with the data published in GCN Observational reports.

\section{\bf Observations and Data analysis }

The CCD photometric monitoring of the optical transient was carried out for 3 nights during 
January 1999 from 23 to 25 using 104-cm Sampurnanand telescope located at 
UPSO, Nainital. The optical transient could be monitored typically for 
about 3 hours in a night. One pixel of the 1024 $\times$ 1024 size CCD chip 
mounted at the f/13 Cassegrain focus of the telescope corresponds to 0.38 
arcsec, and the entire chip covers a field of $\sim$ 6$'\times$6$'$ on the sky.
Except for the night of 23/24 January 1999, others were of photometric
quality. Exposures upto maximum of 20 minutes were generally given. In order 
to improve the S/N ratio of the optical transient, the data have been binned 
in $2 \times 2$ pixel$^2$ and also two or more images of a filter have been 
stacked after correcting them for bias, non-uniformity in the pixels and cosmic 
ray events. Exposure times for the stacked images were generally more than 40
minutes.

As the optical transient was generally quite faint on the stacked images, 
DAOPHOT profile-fitting technique was used for the magnitude determination. 
These magnitudes were calibrated using differential photometric techniques
and standards given by Nilakshi et al. (1999). The results derived in this
way are given in Table 1. One can note that the good photometric sky 
conditions at Nainital enabled us to observe the optical transient till it 
was as faint as $V \sim 22$ mag with an accuracy of 0.10 mag using a 
telescope of aperture 1 m.

\vspace{0.5cm}
\noindent {\bf Table 1.}~$BVR$ CCD magnitudes of the optical transient of 
GRB 990123. Errors are primarily based on the S/N ratio.
\begin{table}[ht]
\begin{tabular}{cclc} \hline 
Time in UT &Filter &\multicolumn{1}{c}{Exposure in minutes}& Magnitude   \\  \hline 
Jan 99  23.958 & $B$ & $30\times 1, 20 \times 1$ & 20.31$\pm$0.08 \\
Jan 99  24.000 & $R$ & $20\times 2$ & 19.61$\pm$0.03 \\
Jan 99  24.940 & $R$ & $20\times 3$ & 20.83$\pm$0.05 \\
Jan 99  24.987 & $B$ & $20\times 3$ & 21.71$\pm$0.11 \\
Jan 99  25.028 & $V$ &  $20\times 1$ &21.08$\pm$0.15 \\
Jan 99  25.940 & $R$ &  $20\times 2$ &21.34$\pm$0.10 \\
Jan 99  25.990 & $V$ &  $20\times 3$ &21.89$\pm$0.10 \\ \hline
\end{tabular}
\end{table}

\section{Results and Discussions}
 Table 1 shows the results of our photometric observations in Johnson $BV$ 
and Cousins $R$ passband. In order to study the photometric light curve of the
optical transient of GRB 990123 we have combined our measurements with the data 
published by various groups in the GCN Observational reports. All photometric 
measurements have been put on the same calibration scale, using the magnitudes 
of the reference stars given by Nilakshi et al. (1999). In this way we have been 
able to take care of the errors arising due to different calibrations. Fig 1 
shows the light curve in $V$, $B$ and $R$  photometric passbands.  
The UPSO observations have been identified
in these plots. The X-axis is log ($t-t_0$) where $t$ is the time of the 
observation and $t_0$ is the time of the trigger of GRB event. All times are 
measured in units of days. The light curves have not yet flattened.

\begin{figure}
\begin{center} \hspace*{-1.5cm}
{\mbox{\psboxto(0cm;17cm){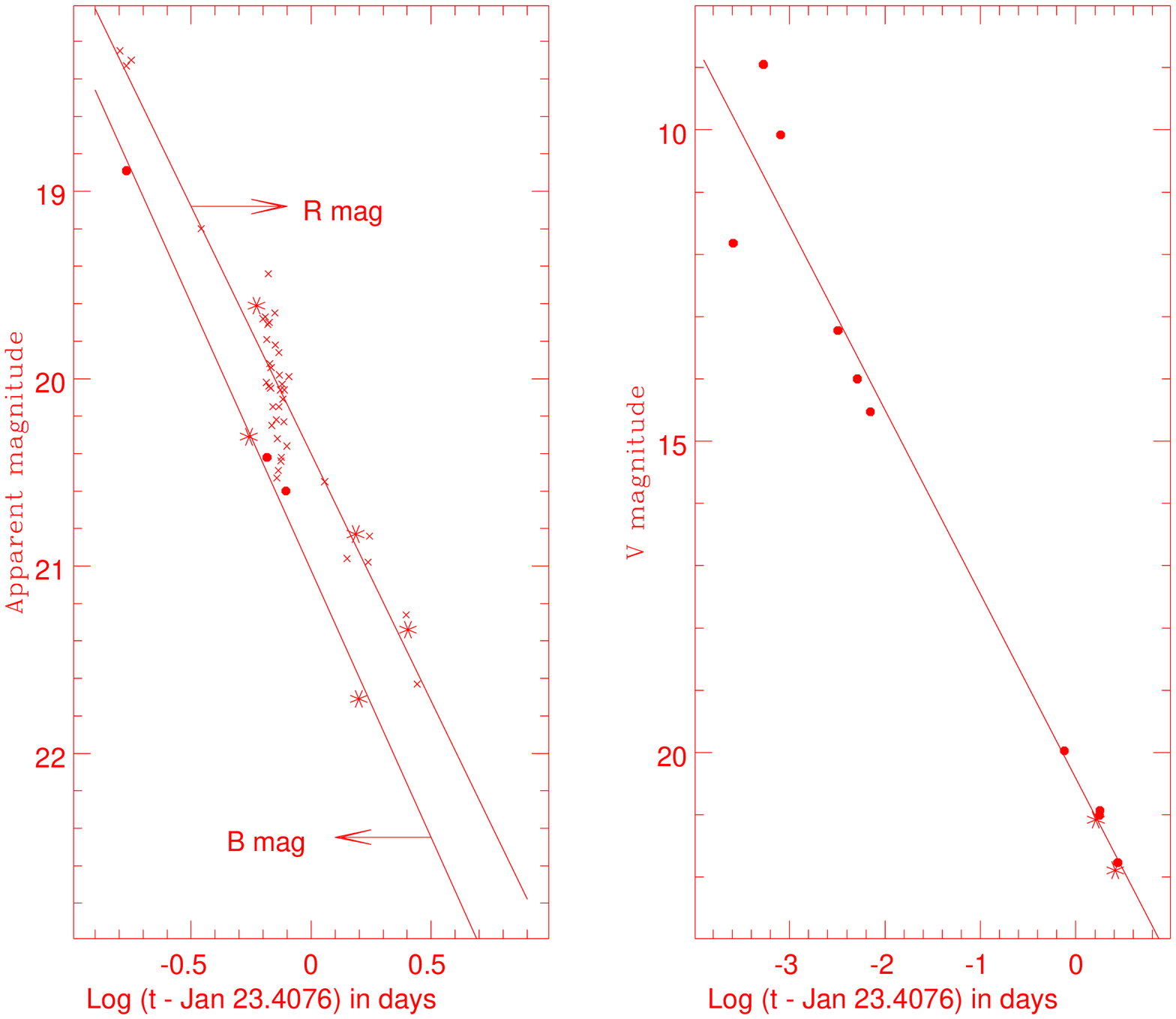}}}
\end{center}
\noindent {\bf Fig. 1. } { $BVR$ photometric light curves of
 the optical transient of GRB 990123. Starred points are from UPSO, Nainital.
Other points are taken from GCN Observational reports (see text). Crosses and 
filled circles in the left panel of the diagram are $R$ and $B$ magnitudes
respectively while $V$ magnitudes are shown as filled circles in the right 
panel of the diagram. }
\end{figure}

The decay of previous optical transients associated with GRBs appears to be well characterized
by a power law $ F(t) \propto (t-t_0)^{-\alpha}$, where $F(t)$ is the flux 
of the optical transient at time $t$ in days and $\alpha$ is the decay 
constant.  Assuming this parametric form and by fitting least square linear 
regressions to the observed magnitudes as function of time, we obtained 
following relations for the $B,V$ and $R$ magnitudes\\
 $ B(t) = (21.02\pm 0.12) + (2.8489\pm 0.17) log (t-t_0) $ \\
 $ V(t) = (20.44\pm 0.18) + (2.8314\pm 0.05) log (t-t_0) $ \\
 $ R(t) = (20.40\pm 0.22) + (2.6487\pm 0.13) log (t-t_0) $ \\
These lines are also plotted in Fig. 1. It can be seen that the UPSO 
observations, within $1-\sigma$ errors, follow the above linear relations. 
The $B$ and $R$ light curves are obtained only during fading phase of the optical
transient, while the $V$ light curve contains both brightening and decaying phase 
of the GRB. In fact, the brightest three points are during the trigger
phase, within 35 seconds after the start of the event. We have therefore
excluded these three points while deriving the decay relation in $V$. However,
their inclusion will not change the value of decay constant significantly. 
The correlation coefficients of these linear relations are always
greater than 0.9. This indicates that assumption of power-law decay for the optical
flux of the optical transient associated with GRB 990123 is justified. Allowing for the factor $-2.5$ involved in converting flux to the magnitude scale, the values 
of flux decay constants are $1.14\pm 0.07, 1.13\pm 0.02$ and $1.06\pm 0.05$
for the $B, V$ and $R$ passbands respectively. In K-passband, the value of
decay constant has been found to be $1.14\pm 0.08$ by Bloom et al. (1999).
Thus, we conclude that the flux decay constants are independent of 
wavelengths  at least in the range of 0.4 to 2.5 micron.
This is in agreement with the decays noted
in earlier optical transients. These decay rates are remarkably similar to those
reported for previous optical transients (see Kulkarni et al. 1998). Such decays are quite
consistent with the synchrotron emission models.

\section{Conclusions}

In gamma-rays the GRB 990123 was one of the top 2 percent of
bursts. In the optical it reached 9th magnitude during the burst itself. 
These properties of the GRB 990123 have led to the suspicion 
that this event was extra strong because of gravitational lensing. A 
consequence of this lensing hypothesis is image splitting. The same
burst would arrive at different times,  with the time difference
proportional to the image separation (e.g., Turner 1999).
One would  therefore expect a repeat of the GRB and
its afterglow. The gamma-ray burst may easily be missed (earth
occultation etc) but its afterglow would be visible for a few days.
We are therefore continuing our observations of the optical transient in $R$ passband.

\bigskip

\noindent {\bf Acknowledgements} The authors are indebted to all those 
who helped in observations. The useful comments given by the referees are
gratefully acknowledged.

\bigskip
\noindent {\bf References} \\
\begin{itemize}
\item [] Alkerlof C.W., McKay T.A., 1999 GCN Observational Report No. 205.
\item [] Anderson M. et al., 1999 in prepration.
\item [] Bloom J.S. et al., 1999 GCN Observational Report No. 240.
\item [] Djorgovski S.G. et al., 1999 GCN Observational Report No. 216. 
\item [] Hjorth J. et al., 1999 GCN Observational Report No. 219.  
\item [] Kelson, D.D. et al., 1999 IAU Circular No. 7096.
\item [] Kulkarni S.R. et al., 1998, Nature, 393, 35.
\item [] Nilakshi, R.K.S. Yadav, V. Mohan, A.K. Pandey, R. Sagar, 1999
   Bull. Astron. Soc. India {\bf 27} (accepted).
\item [] Odewahn S.C., Bloom J.S., Kulkarni S.R., 1999 GCN Observational Report No. 201.
\item [] Piro L. et al., 1999 GCN Observational Report No. 199.
\item [] Turner E. L., 1999 GCN Observational Report No. 221. 
\end{itemize}

\bigskip

\end{document}